\documentclass[twocolumn,showpacs,amsfonts,amsmath,amssymb,aps,pra]{revtex4}
\usepackage{graphicx}
\usepackage{subfigure}
\usepackage{psfrag}

\begin{document}

\title{Exact dynamics and decoherence of two cold bosons in a 1D harmonic trap}
\author{Tomasz Sowi\'nski$^{1,4}$, Miros\l aw Brewczyk$^2$, Mariusz Gajda$^{1,5}$, Kazimierz Rz\c a\.zewski$^{3,5}$}
\affiliation{$^1$Instytut Fizyki PAN, Al. Lotnik\'ow 32/46, 02-668 Warszawa, Poland\\
$^2$Wydzia{\l } Fizyki, Uniwersytet w Bia{\l }ymstoku, ul. Lipowa 41, 15-424 Bia\l ystok, Poland\\
$^3$Centrum Fizyki Teoretycznej PAN, Al. Lotnik\'ow 32/46, 02-668 Warszawa, Poland \\
$^4$Wydzia{\l }  Biologii i Nauk o \'Srodowisku UKSW, ul. W\'oycickiego 1/3 01-938 Warszawa, Poland\\
$^5$Wydzia{\l } Matematyczno-Przyrodniczy SNS UKSW, Al. Lotnik\'ow 32/46, 02-668 Warszawa, Poland}
\author{}

\begin{abstract}
We study dynamics of two interacting ultra cold Bose atoms in a harmonic oscillator potential in one spatial dimension. Making use of the exact solution of the eigenvalue problem of a particle in the delta-like potential we study time evolution of initially separable state of two particles.  The corresponding time dependent single particle density matrix is obtained and diagonalized and single particle orbitals are found. This allows to study decoherence as well as creation of entanglement during the dynamics. The evolution of the orbital corresponding to the largest eigenvalue is then compared to the evolution according to the Gross-Pitaevskii equation. We show that if initially the center of mass and relative degrees of freedom are entangled then the Gross-Pitaevskii equation fails to reproduce the exact dynamics and entanglement is produced dynamically. We stress that predictions of our study can be verified experimentally in an optical lattice in the low-tunneling limit. 
\end{abstract}
  

\maketitle

\section{Introduction}
Theoretical description of a Bose-Einstein condensate of trapped weakly interacting atomic system is traditionally based on a mean field approximation \cite{GPS}. By assuming that many-body wave function can be written in a form of N-fold product state, i.e. that all atoms occupy the same single particle orbital, the stationary Gross-Pitaevskii (GP) equation for the order parameter is found. Assuming further that the N-fold product approximation holds also in dynamical situations one arrives at the time dependent Gross-Pitaevskii equation. Under most of experimental conditions there are no strong correlations in the system and the GP equation turned-out to be extremely fruitful in predicting and describing a variety of features of those systems. Soon it occurred that also high energy solutions of the GP equation can be useful in studying Bose systems at finite temperatures. The GP equation has become a work horse of the theory of weakly interacting ultra cold bosons. 

On the other hand examples when the mean field description does not reproduce the real dynamics have been studied. For instance direct comparison of the mean field and many body theory of vortex nucleation showed that the GP equation fails to describe this phenomenon \cite{Maciek1,Maciek2,Maciek3}. Similarly a mean field description of attractive Bose systems encounters difficulties \cite{Gajda,Gajda2,Hedelberg}. Due to permanent progress in experimental techniques the physics of ultra cold atomic gases started to penetrate areas traditionally associated with condensed matter physics where correlations play a crucial role. Evidently, in such situations simple mean field description based on the GP equation becomes insufficient. The Mott insulator-superfluid transition \cite{Bloch} or the Tonks-Girardeau gas \cite{Tonks,Tonks2} are some examples. 

It is commonly believed that creation of the Mott insulator with a small and controlled number of atoms per lattice site allows for applications of such systems in quantum information. All quantum information processing is based on the quantum correlations which cannot be described by any classical theory based on local realism. States showing these non-local correlations are known as entangled states. Two entangled spin--${1\over2}$ particles are the essence of the Einstein, Podolsky, and Rosen paradox \cite{EPR}. The entangled states vialote Bell inequalities \cite{Bell}. It has been realized that also continuous variable entangled systems can be used in quantum computations \cite{qcomp1,qcomp2,qcomp3,qcomp4,qcomp5,qcomp6,qcomp7}. Several authors \cite{Entropy1,Entropy2,Entropy3} studied recently creation of entanglement during dynamics of two interacting particles in a harmonic trap. In particular it has been shown that this dynamically entangled state violates a Bell-type inequality for a certain choice of observables \cite{Entropy1}. 

In this paper we study dynamically created entanglement, and its measure -- the von Neuman entropy, for a realistic system of two identical bosonic atoms in a harmonic trap. We consider low energy collisions of the atoms. At such energies the range of van der Waals interactions is smaller than the s-wave scattering length. Therefore, the interaction potential can be approximated by a  contact pseudo potential. This approximation occurred to be in excellent agreement with experimental results \cite{Esslinger} where binding energy of molecular system has been measured. The molecules were created from atoms in an optical lattice in the limit of small tunneling. This experimental arrangement is perfectly suited for a study of exact dynamics of two trapped atoms. We consider a realistic case of two atoms per lattice site deep in the Mott insulator phase. By applying a Bragg pulses \cite{Phillips} one creates a state in which each atom is in a superposition of two counter propagating wave packets. Initially, such a state is a two-fold product state of two identical wave functions, i.e. is separable. Each wave function has two components moving initially with opposite momenta. The center of mass dynamics is generated by a different Hamiltonian than dynamics of the relative coordinate therefore this two particle continuous variable system becomes entangled. We use the von Neuman entropy as a measure of the entanglement and study its behavior in time for various interaction strengths. We also analyze a coherence of the system which is directly related to the largest eigenvalue of the single particle density matrix and compare the exact dynamics to the mean field description based on the GP equation.             

\section{Two bosons in a harmonic trap}
We are going to study dynamics of the simplest nontrivial system -- two atoms confined in a one dimensional harmonic potential. In fact generalization of our results to two or three spatial dimensions is straightforward. We limit our analysis to the 1D case as this situation captures all features of the dynamics. For simplicity we are using harmonic-oscillator units. It means that all energies are measured in $\hbar \omega$, all lengths in $\sqrt{\hbar/m\omega}$, and all momenta in $\sqrt{\hbar m \omega}$. Hamiltonian of the system of two interacting bosons in the harmonic trap has the form:
\begin{equation} \label{Hamiltonian}
{\cal H} = -\frac{1}{2}\frac{\partial^2}{\partial x_1^2} -\frac{1}{2}\frac{\partial^2}{\partial x_2^2}+\frac{1}{2}\left(x_1^2+x_2^2\right) + g \delta(x_1-x_2)
\end{equation}
where $x_1$ and $x_2$ are positions of atoms interacting via a short range potential modeled by the delta function. This form of the short range interaction is justified in the limit of vanishing relative velocity of colliding atoms, where atomic de Broglie wavelength is much larger then a range of two body potential. In 2D and 3D the corresponding Hamiltonian is not a self adjoint operator. To correct for this fact a regularization is required. In contrast to many dimensions, the regularization of the delta function is not necessary  in one dimensional case \cite{Se86}. In 1D the parameter $g$ is given by $g=-2/a_0$, where $a_0$ is a scattering length \cite{Al88}. It is worth to notice that finite range interactions between particles modeled by the Gaussian function in the context of the dynamics of two bosons was studied in \cite{Sascha}.

To demonstrate entanglement formation we study the evolution of two bosons which initially are in a product quantum state
\begin{equation}
\Psi_0(x_1,x_2) = \Phi_0(x_1)\Phi_0(x_2).
\end{equation}
Function $\Phi_0(x)$ is a one-particle wave function called the order parameter in the mean field context. 

The exact dynamics of the two interacting bosons in the harmonic trap can be found because all eigenstates of the full two-body Hamiltonian \eqref{Hamiltonian} are known. They are found in \cite{Bu98}. The two particle problem has to be first brought to a single particle one by introducing the center of mass and the relative coordinates:
\begin{subequations} \label{relations}
\begin{align}
X &= \frac{1}{\sqrt{2}}(x_1+x_2) \\
\xi &= \frac{1}{\sqrt{2}}(x_1-x_2) 
\end{align} 
\end{subequations}
In these coordinates Hamiltonian \eqref{Hamiltonian} separates into two independent parts -- the center of mass part, and the relative part:
\begin{subequations}
\begin{align}
{\cal H}_{\mathtt{CM}} &= -\frac{1}{2}\frac{\mathrm{d}^2}{\mathrm{d}X^2}+\frac{1}{2}X^2 \label{HamiltonianCM}\\
{\cal H}_{\mathtt{REL}} &= -\frac{1}{2}\frac{\mathrm{d}^2}{\mathrm{d}\xi^2}+\frac{1}{2}\xi^2 +\frac{\sqrt{2}}{2}g\,\delta(\xi) \label{HamiltonianRel}
\end{align}
\end{subequations}
As one can see, the dynamics of the center of mass is described by the standard one dimensional harmonic oscillator Hamiltonian \eqref{HamiltonianCM}. Its eigenstates are well known and have a standard form
\begin{subequations}
\begin{equation}
\chi_n(X) = \frac{\pi^{-1/4}}{\sqrt{2^n\,n!}}\,\mathrm{H}_n(X)\,\mathrm{e}^{-X^2/2},
\end{equation}
where $\mathrm{H}_n(x)$ are Hermite polynomials. The energy of $n$-th eigenstate in our units is obviously given by
\begin{equation}
{\cal E}_n = n+\frac{1}{2}
\end{equation}
\end{subequations}
The eigenstates of the Hamiltonian \eqref{HamiltonianRel} describing relative dynamics of two particles are given in \cite{Bu98} and for one dimensional problem have a form
\begin{subequations}
\begin{align}
\varphi_{m}(\xi) &= \frac{\pi^{-1/4}}{\sqrt{2^m\,m!}}\,\mathrm{H}_m(\xi)\,\mathrm{e}^{-\xi^2/2}, & m\,\, \textrm{odd} \\
\varphi_{m}(\xi) &={\cal N}_m\,\mathrm{U}(-\nu_m,\frac{1}{2},\xi^2)\,\mathrm{e}^{-\xi^2/2},  & m\,\, \textrm{even} 
\end{align}
\end{subequations}
where 
$\mathrm{U}(\alpha,\beta,x)$ are confluent hypergeometric functions, and ${\cal N}_m$ are normalization coefficients. Since the wave function of identical bosons must be symmetric under exchange of the two particles, therefore the physical wave function is composed from functions with even $m$ only. The energies $E_m$ of these even states are given by a sequence of zeros of the function:
\begin{align}
f(E)=\frac{\Gamma(-E/2+3/4)}{\Gamma(-E/2+1/4)}-\frac{1}{a_0}.
\end{align}
The quantum number $\nu$ is equal to $\nu_m=(2E_m-1)/4$.
The initial wave function can be easily decomposed to the superposition of the eigenstates of the Hamiltonian:
\begin{equation} \label{Decompos}
\Psi_0(\xi,X) = \sum_{nm} \alpha_{nm}\, \chi_n(X)\varphi_m(\xi)
\end{equation}
Obviously the evolution of the initial two boson state is given by:
\begin{equation}
\Psi(\xi,X,t) = \sum_{nm} \alpha_{nm}\, \chi_n(X)\varphi_m(\xi)\,\mathrm{e}^{-i({\cal E}_n+E_m)t}.
\end{equation}
The last step is to return to the original coordinates by using relations \eqref{relations}: 
\begin{multline}
\Psi(x_1,x_2,t) = \sum_{nm} \alpha_{nm}\, \chi_n\left(\frac{x_1+x_2}{\sqrt{2}}\right)\\ \times \varphi_m\left(\frac{x_1-x_2}{\sqrt{2}}\right) \mathrm{e}^{-i({\cal E}_n+E_m)t}
\label{w_fun}
\end{multline}

Standard method of detection of ultra cold trapped atomic systems are destructive. The optical lattice potential is turned off and the system is allowed to expand balistically. Only after expansion a size of the system exceeds a resolution of a CCD camera. The picture of the CCD camera gives therefore direct insight into the initial momentum distribution of atoms. The wave function Eq.(\ref{w_fun}) written in the momentum space of the two atoms is:
\begin{equation}
\psi(k_1,k_2,t)=\int_{-\infty}^\infty \mathrm{d}x_1 \int_{-\infty}^\infty \mathrm{d}x_2 e^{-i k_1 x_1} e^{-i k_2 x_2} \Psi(x_1,x_2,t).
\end{equation} 
In repeated single particle detections preceded by the ballistic expansion of the system one-particle momentum distribution is monitored:
\begin{equation} \label{densityExact}
n_{\mathtt{Exact}}(k,t)=\rho_1(k,k,t),
\end{equation}
where $\rho_1(k,k',t)$ is the reduced one particle density matrix in the momentum representation:
\begin{equation} \label{DensityMatrix}
\rho_1(k,k',t) = \int_{-\infty}^\infty\!\!\mathrm{d}k_2\,\psi^*(k,k_2,t)\psi(k',k_2,t)
\end{equation}
By making its spectral decomposition we can determine the number of orbitals and their relative occupations needed for accurate description of the two bosons dynamics. Time dependence of the eigenvalues of the density matrix is discussed below. Let us mention that the largest eigenvalue is a direct measure of the coherence of the system.

We shall compare this exact dynamics with the approximate one governed by the Gross-Pitaevskii equation. The main idea leading to the mean field approximation relies on the assumption that generation of entanglement between bosons during the evolution is negligible and therefore the quantum state of the system remains separable. In other words all correlations between bosons are neglected and the same wave functions of every particle is assumed during the evolution:
\begin{equation}
\Psi(x_1,x_2,t) = \Phi(x_1,t)\Phi(x_2,t).
\end{equation}
This assumption leads directly to the Gross-Pitaevskii equation which determines the dynamics of the one-particle wave function $\Phi(x,t)$:
\begin{equation}
i\partial _t \Phi(x,t) = \left(-\frac{1}{2}\frac{\partial^2}{\partial x^2}+\frac{1}{2}x^2+g|\Phi(x,t)|^2\right)\Phi(x,t).
\end{equation}
The probability density in momentum space reads:
\begin{equation} \label{densityGP}
n_{\mathtt{GP}}(k,t)=|\phi(k,t)|^2,
\end{equation}
where $\phi(k,t)$ is the Fourier transform of the time dependent solution of the GP equation, $\phi(k,t)=\int {\rm d}x\,e^{-ikx} \Phi(x,t)$. We compare the exact one-particle momentum distribution with that predicted by the Gross-Pitaevskii approximation \eqref{densityGP}.  Moreover, in the situation when many eigenvalues of the density matrix \eqref{DensityMatrix} are of the same order we can also compare the Gross-Pitaevskii momentum distribution \eqref{densityGP} with the momentum distribution of the dominant orbital obtained from diagonalization of the exact one-particle density matrix in the momentum space. Obviously, in a general case the GP solution overestimates the coherence of the system.

The GP  equation is solved numerically on a spatial grid of $N_p=2^{10}$ points separated by $\delta x = 5\cdot 10^{-2}$. The time step is equal to $\delta t = \pi/4\cdot 10^{-3}$. We use the split-operator method which is very stable for the chosen temporal and spatial steps.

\section{Results}

To make the detailed comparison we concentrate on a one particular class of the initial states. We assume that at the beginning each particle is in the state described by the Schr{\"o}dinger cat like wave function
\begin{equation}
\Phi_0(x) = {\cal N} \left[\mathrm{e}^{-(x-L)^2/2}+\mathrm{e}^{-(x+L)^2/2}\right]
\end{equation}
Parameter $L$ measures the separation between two wave packets moving in the opposite direction in the relative coordinates space. Such a choice is motivated by the preparation procedure described above, i.e. we assume that Bragg pulses bringing the atoms into the superposition of wave packets moving in opposite directions are applied. When $L=0$ then the initial state is very close to the ground state of the system so we expect that the exact dynamics is almost indistinguishable from the dynamics in the mean filed approximation.  For large $L$ the initial state is still separable but it is highly delocalized. Relative and center of mass degrees of freedom are entangled in the initial state. They evolve in a different way, therefore we expect that the exact dynamics could be dramatically different than the dynamics predicted by a simple mean field approach. 

\subsection{Dependence on delocalization of one-particle state}

\begin{figure}[ht]
\centering
\includegraphics[scale=0.6]{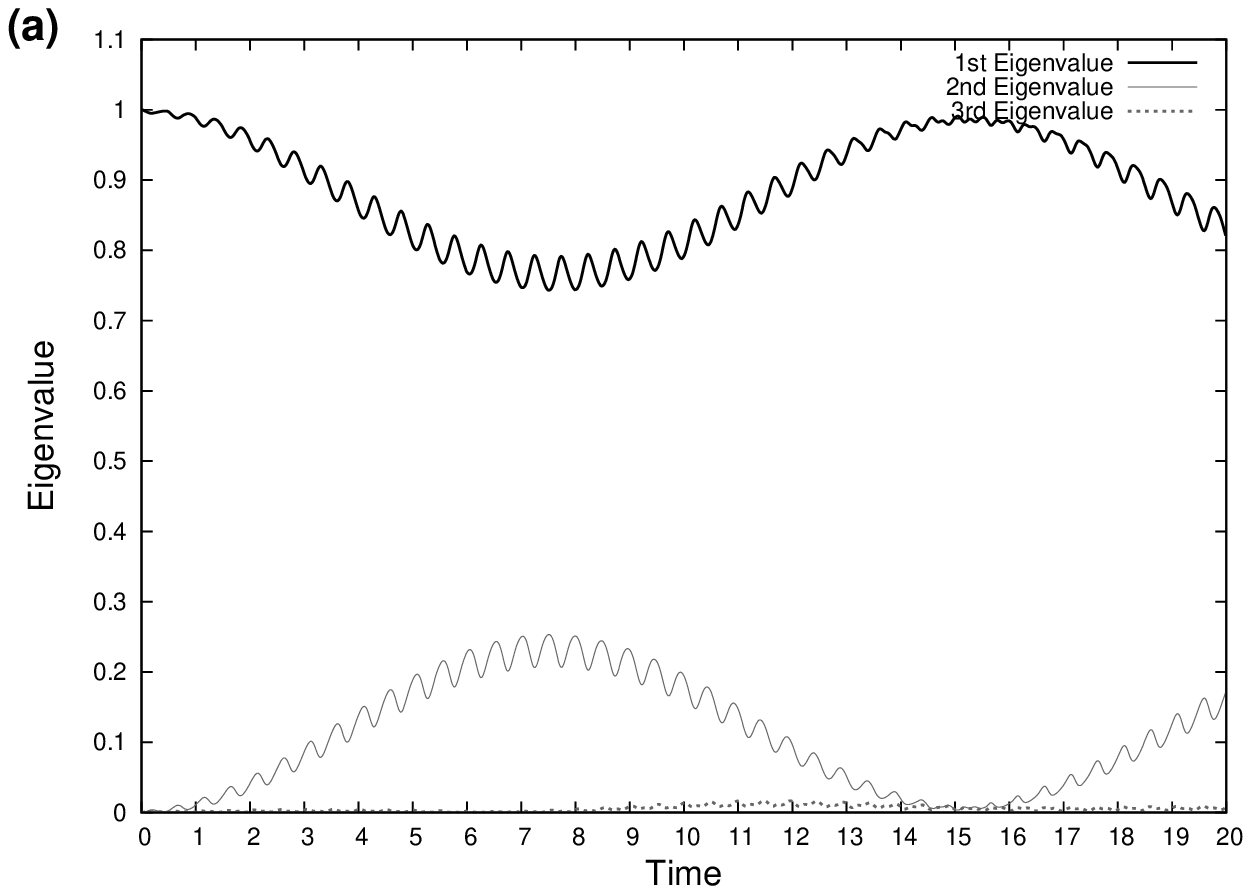}
\includegraphics[scale=0.9]{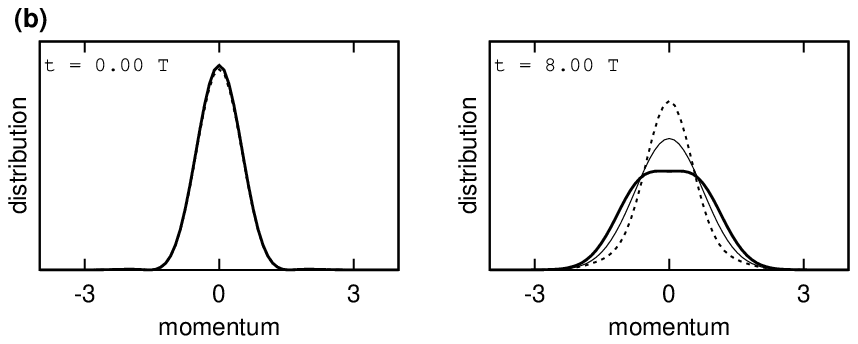}
\caption{(a) Eigenvalues of the one-particle density matrix \eqref{DensityMatrix}. Unit of time is equal to the period of the trap. In this situation (parameters: $g=-0.2$, $L=1$) the initial state is not far from the ground state of the system. One eigenvalue still dominates, therefore system should be quite well described by the mean field approximation. (b) Two plots present the one-particle momentum distributions predicted by the exact (thick solid line) and the Gross-Pitaevskii solutions (dotted line) in two interesting moments. Third (thin solid) line comes from the exact solution and presents the momentum distribution of the first orbital.  As was expected all three predictions are almost the same for considered set of parameters. Movie presenting time evolution of momentum distributions is available on-line \cite{movies}}
\label{fig:fig1}
\end{figure}

First let us discuss situation for generic interaction strength $g=-0.2$ ($a=10$) when $L=1$, i.e. when the extension  of the initial state is equal to the trap length unit. We observe that the single particle density matrix obtained from the exact dynamics develops more then one nonzero eigenvalue, i.e. many one particle orbitals are involved. Fig. \ref{fig:fig1}a shows time dependence of the eigenvalues of the one-particle density matrix \eqref{DensityMatrix}. Because one of the eigenvalues is incessantly much larger than the others the system coherence is large and the Gross-Pitaevskii description is quite accurate in this case. Time dependence of the momentum distributions deduced from the Gross-Pitaevskii equation and the exact solution are shown in Fig. \ref{fig:fig1}b and they are in agreement with our predictions (whole time dependence of momentum distributions is available on-line \cite{movies}).

\begin{figure}[ht]
\centering
\includegraphics[scale=0.6]{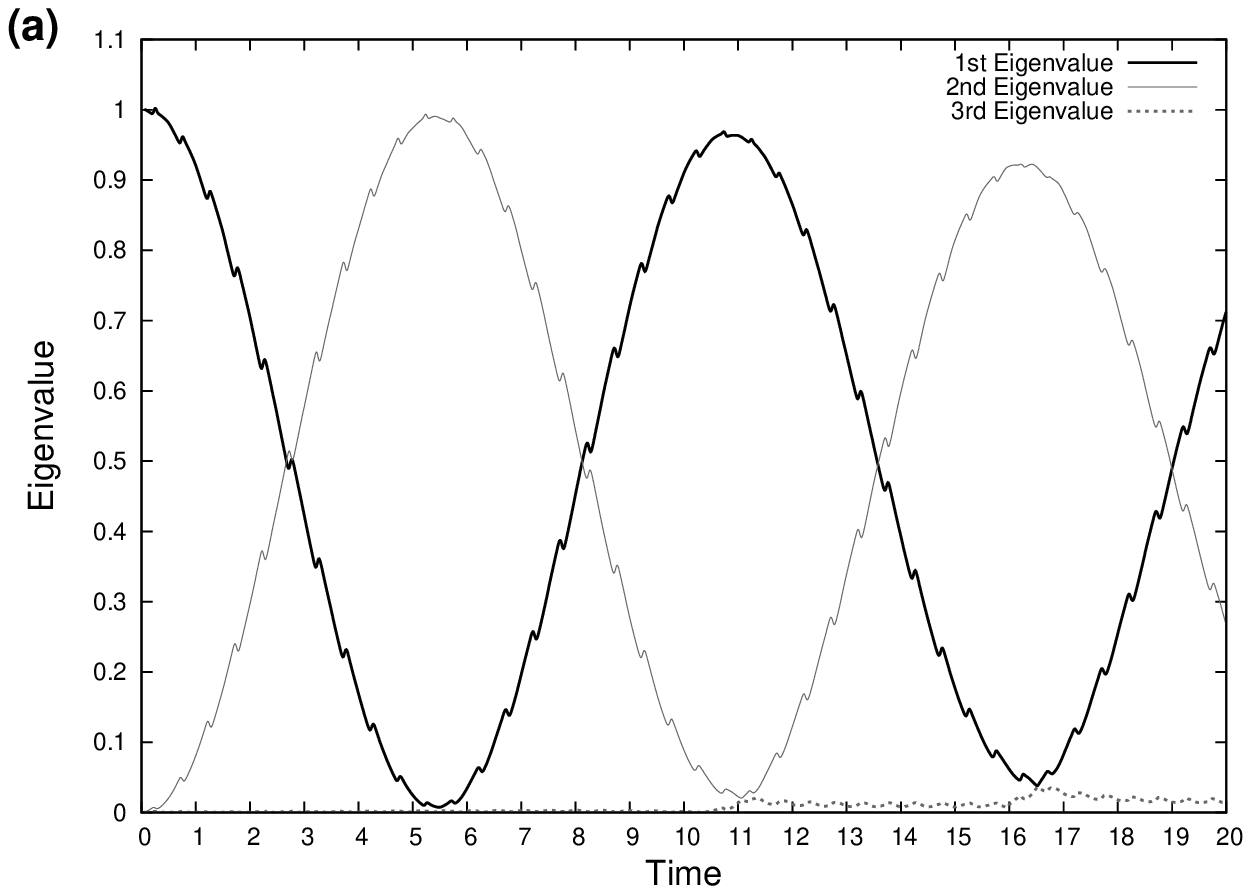}
\includegraphics[scale=0.9]{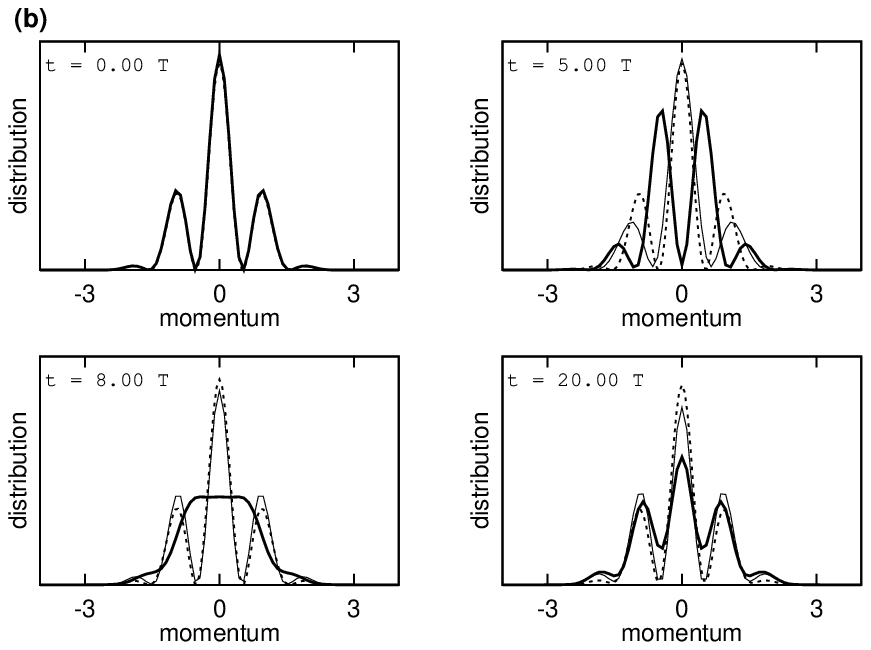}
\caption{(a) Eigenvalues of the one-particle density matrix \eqref{DensityMatrix} for $g=-0.2$, $L=3$. Unit of time is equal to the period of the trap. In this situation the initial state is a product of highly delocalized one-particle wave functions. There is no one dominant eigenvalue during the evolution and therefore the Gross-Pitaevskii equation will not predict dynamics correctly. (b) Time dependence of the one-particle momentum distribution predicted by the exact (thick solid line) and Gross-Pitaevskii solutions (doted line). As long as the first eigenvalue dominates during the time evolution the predictions are almost the same. After five periods the second eigenvalue is the largest one and therefore the predictions are highly different. Solutions of the exact and GP dynamics become similar after eleven trap periods when the first eigenvalue starts to dominate again. Notice that third (thin solid) line presenting momentum distribution of first one-particle orbital of an exact solution recovers predictions of the Gross-Pitaevskii equation. The movie is available on-line \cite{movies}. }
\label{fig:fig2}
\end{figure}

Situation changes dramatically when we increase the delocalization parameter. When $L$ is large enough then a few orbitals can play the crucial role in the dynamics and the mean field approximation is no longer valid. Fig. \ref{fig:fig2}a shows the time dependence of the eigenvalues of the density matrix for $L=3$. As we see, the main orbital (its eigenvalue is represented by a thick solid line) initially dominates. But after a few periods of the trap oscillations the other orbital becomes much more important than the first one. The dynamics is obviously much more complicated than it is predicted by the mean field approach. It is clear when we compare the momentum density distribution predicted by the exact and the mean field solutions (Fig. \ref{fig:fig2}b and movie available on-line \cite{movies}).

We see that evidently Gross-Pitaevskii equation properly describes the dynamics of the first orbital rather then the whole system, Fig. \ref{fig:fig2}b. It is the reason why the Gross-Pitaevskii equation gives good predictions when only one eigenvalue of the one particle matrix dominates during the entire evolution.

\begin{figure} 
\centering
\includegraphics[scale=0.9]{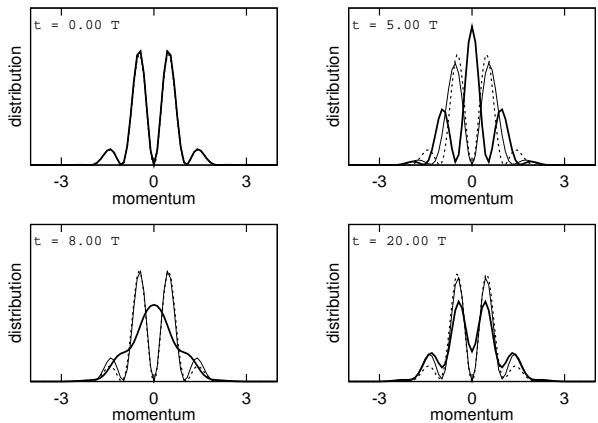}
\caption{Time dependence of the one-particle momentum distribution for the antisymmetric initial state $\Phi_0(x) = {\cal N} \left[\mathrm{e}^{-(x-L)^2/2}-\mathrm{e}^{-(x+L)^2/2}\right]$ with $L=3$. Thick solid line represents the density predicted by an exact solution, while doted one the density coming from the mean field approach. Properties of the Gross-Pitaevskii equation provide that if the function $\Phi_0(x)$ is antisymmetric under $x\rightarrow -x$ symmetry then it will stay antisymmetric during the whole evolution. It is not true for the one-particle density predicted by an exact solution. Thin solid line presents momentum distribution of the first one-particle orbital of the exact solution. As we see its spatial reversal symmetry is preserved during the evolution. It shows once more that Gross-Pitaevskii equation describes properly the dynamics of the first orbital only. Whole movie is available on-line \cite{movies}.}
\label{fig:fig5}
\end{figure}

It is also interesting to study similarities and differences between prediction of the mean field approach and exact solution in the situation when the initial state of each particle is antisymmetric in position space, i.e. is described by the wave function of the form
\begin{equation}
\Phi_0(x) = {\cal N} \left[\mathrm{e}^{-(x-L)^2/2}-\mathrm{e}^{-(x+L)^2/2}\right].
\end{equation}
Nevertheless the wave function of the system is still symmetric under particle exchange. Since corresponding GP Hamiltonian is invariant under  reflection $x \rightarrow -x$, therefore the symmetry of the initial state will be preserved. As we observe, it is not true for the exact two body dynamics. The evolution preserves only the symmetry of each orbital separately, but not the symmetry of the whole system. It is clearly demonstrated in Fig. \ref{fig:fig5} where we compare the momentum distribution predicted by the mean field approach with the single particle density obtained from the exact dynamics. Time dependence of the eigenvalues of one-particle density matrix is identical as the one for the corresponding symmetric case (Fig. \ref{fig:fig2}a). Whole movie is also available on-line \cite{movies}.

\subsection{Dependence on interaction strength}

Now we want to show that correctness of the mean field approximation significantly depends on the interaction strength parameter $g$. It is quite obvious that in the situation when the interaction is switched off, the two particles initially in the state which is not entangled (product state) will stay in such a state during the whole evolution even for a highly delocalized state. In this case the mean field approximation naturally leads to the same solution as the exact solution. It is the interparticle interaction which can produce entangled two body states during the evolution.  

\begin{figure}
\centering 
\includegraphics[scale=0.6]{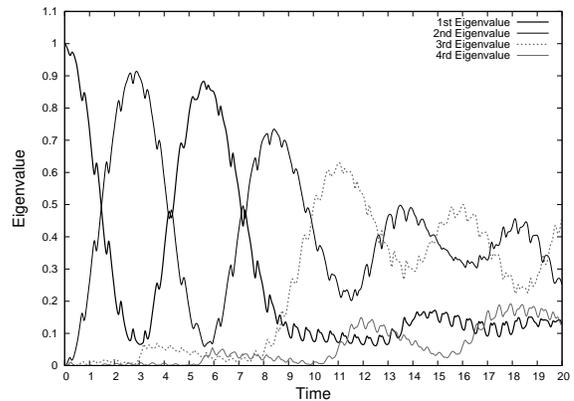}
\caption{Eigenvalues of the one-particle density matrix as functions of time for parameters: $g=-0.4$, $L=2$. The interaction between bosons is strong and the initial state of one particle is highly delocalized. In such a situation many orbitals play a crucial role during the evolution of the system. Therefore the exact dynamics is much more complicated that the dynamics predicted by the mean field approximation. }
\label{fig:fig3}
\end{figure}

Time dependence of the eigenvalues for a moderate interaction strength  ($g=-0.2$) is presented in Fig. \ref{fig:fig1} and \ref{fig:fig2}. In those situations only two eigenvalues (i.e. two orbitals) are important for many trap periods. For stronger interactions this picture changes significantly. Time evolution of eigenvalues for strong interaction ($g=-0.4$) and $L=2$ are shown in Fig. \ref{fig:fig3}. After a few trap periods many different orbitals become important. Moreover the orbital which dominates at the beginning becomes unimportant after a very short time. Therefore we do not expect that the Gross-Pitaevskii approximation may give correct predictions in this case.

\begin{figure}
\centering
\includegraphics[scale=0.6]{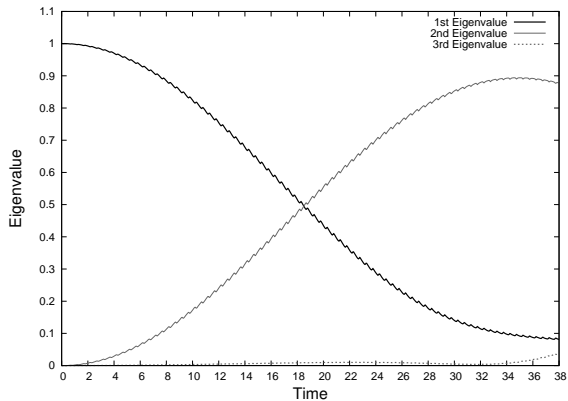}
\caption{Eigenvalues of the one-particle density matrix as functions of time for parameters: $g=-0.04$, $L=2$. In this situation the interaction between bosons is very week but the initial state is far from the ground state of the system. During the first eighteen trap periods only one eigenvalue dominates, therefore the dynamics of the system can be quite correctly described by the mean field approximation for a long time. Notice that time scale is two times larger than in the previous situation.}
\label{fig:fig4}
\end{figure}

On the other hand when the interaction is very weak we can expect that the production of entanglement will be very slow even for highly delocalized states and therefore the mean field approximation may be correct for a long evolution time. Time dependence of the eigenvalues of the one-particle density matrix when the interaction is weak but the initial state is highly delocalized is presented in fig. \ref{fig:fig4}. 

\subsection{Revivals of product states}
\begin{figure}
\centering
\psfrag{Amplitude}{$|\alpha|^2$}
\includegraphics[scale=0.6]{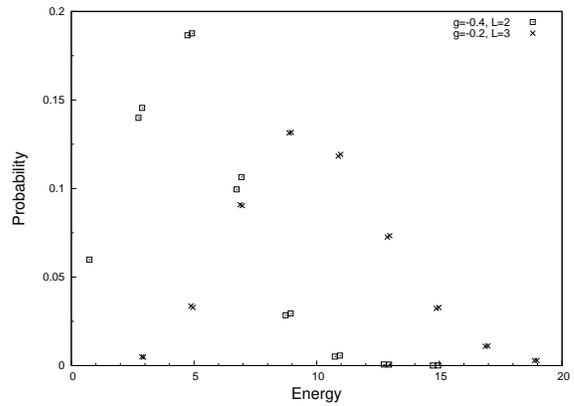}
\caption{Energy spectrum of two initial states described by the parameters: $g=-0.4$ and $L=2$ (crosses); $g=-0.2$ and $L=3$ (squares). Note that both spectra are well picked and energies are `paired'.}
\label{fig:fig8}
\end{figure}

Looking at fig. \ref{fig:fig2} and fig. \ref{fig:fig4} one can observe that as initially only one eigenvalue dominates in the Schmidt decomposition of the single particle density matrix, the other eigenvalues become more important at later times. However, the time dependence of the dominant eigenvalue exhibits some oscillations and a partial revival of the `initial' eigenvalue can be observed. It is interesting to find a physical explanation of this behavior. To this end in fig. \ref{fig:fig8} we show the spectrum of the two-particle state for the two studied parameter sets $g=-0.2$ and $L=3$ (as in fig. \ref{fig:fig2}) and $g=-0.4$ and $L=2$ (as in fig. \ref{fig:fig4}). In this figure we plot the probability of the given eigenenergy, $|\alpha_{nm}|^2$, resulting from the decomposition \eqref{Decompos}. 

First let us notice that the eigenenergies do appear `in pairs'. The effect of `pairing' of eigenenergies can be easily explained. For given $n$ and $m$ the eigenenergy has two components. ${\cal E}_n$ is the energy of the center of mass while $E_m$ corresponds to the relative coordinate. As was mentioned before, energies of the relative excitations are very close to the energies of harmonic trap. This is because the potential in the relative coordinate space is the harmonic potential of the trap modified at $\xi =0$ by the presence of delta function. This delta function shifts the harmonic energy by very small amount. Therefore state labeled by $(n,m)$ is almost degenerate with the state labeled by $(m,n)$ and therefore spectrum is paired. 

If by $\Delta$ we denote the difference between two energies of dominant pair (maximum in the spectral decomposition) it is clear that after time $T_{\mathtt{R}} = 2\pi/ \Delta$ these two the most important amplitudes of the two-atom wavefunction match in phase and the partial revival of the product state can be observed. This is signified by a reappearance of the initial eigenvalue of the single particle matrix. The revival time calculated this way for parameters of   Fig. \ref{fig:fig2} is equal to $T_{\mathtt{R}}=10.96$ while for Fig. \ref{fig:fig3} is $T_{\mathtt{R}}=5.56$ which agrees perfectly with predictions of the exact dynamics. 

Obviously revival time $t_R$ depends on the initial state as well as eigenmodes of the Hamiltionian \eqref{Hamiltonian}. 

In addition to this large time oscillations of the eigenvalues of the density matrix some small fast oscillations can be also observed. This fast modulations appear every half of the trap period when two wavepackets meet at the trap center and results from interaction between them. 

\subsection{Entanglement of particles}
Mutual interactions between particles obviously leads to the quantum correlations between particles. To study them we use the correlation measure introduced in \cite{Kazik}:
\begin{equation} \label{Ent}
\mathbf{K}(\rho_1) = \left(\sum_i \lambda_i^2\right)^{-1}
\end{equation}
where $\lambda_i$ are the eigenvalues of the one-particle density matrix $\rho_1$. This measure has very simple interpretation. It gives an effective number of single particle orbitals occupied in the given many body state. In particular when one-particle density matrix has $n$ equal eigenvalues then $\mathbf{K}=n$. 

\begin{figure}
\centering
\psfrag{kyaxis}{$\mathbf{K}(\rho_1)$}
\psfrag{syaxis}{$\mathbf{S}(\rho_1)$}
\includegraphics[scale=0.6]{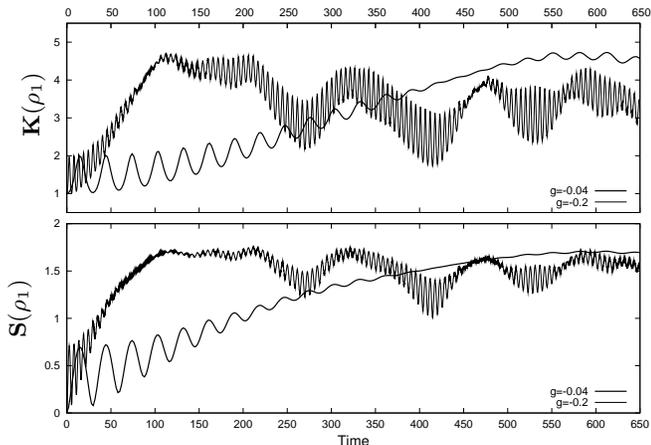}
\caption{Time dependence of the number of dominant eigenvalues $\mathbf{K}$ defined in \eqref{Ent} and of the von Neumann entropy $\mathbf{S}$ defined in \eqref{NeumannEntropy} for $g=-0.04$ (thick line) and $g=-0.2$ (thin line). Other parameters are the same as in Fig. \ref{fig:fig4}.}
\label{fig:measures}
\end{figure}

Other commonly used \cite{Entropy1,Entropy2,Entropy3,Entropy0} measure of entanglement in the system is von Neumann entropy defined as 
\begin{align} \label{NeumannEntropy}
\mathbf{S}(\rho_1) &= -\mathrm{Tr}\left(\rho_1 \mathrm{log}\,\rho_1\right) = -\sum_i\lambda_i\, \mathrm{log}\,\lambda_i
\end{align}
This entropy is even more interesting than the number of dominant eigenvalues $\mathbf{K}$ since it is directly connected with the entropy defined in thermodynamical context. Time dependence of this two measures of entanglement in the system for two different regimes of interaction strength are presented in Fig. \ref{fig:measures}. Obviously in the beginning, when the system is in separable state, entanglement and von Neumann entropy are equal to $1$ and $0$ respectively. We observe that correlation $\mathbf{K}$ and entropy $\mathbf{S}$ increase in time and seem to saturate for large time. Even though they have different physical interpretation they behave very similarly which might seem quite surprising. They reach `stationary regime' faster for stronger interactions. 

Both quantities exhibit fast oscillations modulated by a slowly varying functions. These fast oscillations are related to partial revival of dominant eigenvalue discussed in previous subsection. Every minimum observed in correlation function corresponds to the moment when there is a dominant eigenvalue in the Schmidt decomposition of the one-particle density matrix. Let us remind that this revivals are related to phase matching of two dominant eigenmodes of the two particle state. 

Long time modulations of correlation functions are related to the quantum nature of the system and discreetness of the energy spectrum. In such cases evolution is always quasi-periodic and due to the interference of amplitudes long time scale oscillations do appear. In our case the number of modes with no zero amplitudes is relatively small and therefore oscillations of correlation functions appear on a time scale of few hundred trap periods. 

\section{Summary}
In this paper we study the exact dynamics of two particles trapped in a harmonic trap and interacting by a  contact potential. We assumed that initially each  particle is transferred by the Bragg pulses to the state being the superposition of two wave packets moving in opposite directions. We show that the two particle state, although initially being a product state does not preserve the product form during the evolution. The reason is that the initial state entangles the center of mass and relative coordinates of the two particle system. These two degrees of freedom evolve according to different Hamiltonians. As a result the single particle reduced matrix develops many eigenvalues during the evolution what signifies decreasing coherence of the system. This situation cannot be correctly described by the GP equation. Our predictions can be verified in the experiment with deep optical lattices when two atoms occupy each site. We show one-particle momenta distributions for different initial states and compare them to those obtained from the mean-field dynamics.  The differences between the two signify the two atom entanglement. The momentum distribution is directly measured by exposure of the system to a resonant light after ballistic expansion and  therefore creation of entanglement in the two particle system can be easily traced in time and compared to exact solutions. We monitored the von Neumann entropy which is common measure of entanglement. We show that entanglement is dynamically created during evolution, however it is not very surprising for interacting system. A comment about a system of two fermionic particles would be also in place. As two identical fermions do not interact in the s-wave channel, as long as other partial waves can be neglected, their dynamic is driven by the noninteracting Hamiltonian. On the other hand, if the spatial part of a wave function of two fermions is symmetric and a spin part is responsible for the antisymmetrization of the total wave function, then our exact solution
evidently applies to such a situation.
 
\begin{acknowledgments}
We acknowledge the support of the Polish Ministry of Science and Education grants for 2008-2010 (T.S.), 2009-2011 (M.B., K.R.), and EU project NAME-QUAM (M.G.).
\end{acknowledgments}

\end{document}